\documentclass[twocolumn,showpacs,amsmath,amssymb,pr,superscriptaddress]{revtex4-1}
\usepackage{graphicx}
\usepackage{epsf}
\usepackage{ulem}
\usepackage{color}
\usepackage[unicode=true,colorlinks=true,urlcolor=blue,citecolor=blue]{hyperref}

\newcommand{\resub}[1]{{\color{black} #1}}

\begin{document}

\title{Dresselhaus spin-orbit interaction in the p-AlGaAs/GaAs/AlGaAs structure with a square quantum well: Surface Acoustic Waves Study}

\author{I. L. Drichko}
\author{I. Yu. Smirnov}
\affiliation{Ioffe Physical–Technical Institute, Russian Academy of Sciences, St. Petersburg, 194021 Russia}
\author{A.~V.~Suslov}
\affiliation{National High Magnetic Field Laboratory, Tallahassee, FL 32310, USA}
\author{K.~W.~Baldwin}
\author{L.~N.~Pfeiffer}
\author{K.~W.~West}
\affiliation{Department of Electrical Engineering, Princeton University, Princeton, NJ 08544, USA}

\begin{abstract}

The effect of spin-orbit interaction was studied in a high-quality $p$-AlGaAs/GaAs/AlGaAs structure with a square quantum well using acoustic methods.
The structure grown on a GaAs (100) substrate
was symmetrically doped with carbon on both sides of the quantum well. Shubnikov\resub{--}de Haas-type oscillations of the ac conductance of two-dimensional holes were measured. At a low magnetic field $B <$2~T conductance oscillations undergo beating induced by a spin-orbit interaction. Analysis of the beating character made it possible to separate the conductance contributions from the two heavy holes subbands split by the spin-orbit interaction. For each of the subbands the values of the effective masses and quantum relaxation times have been determined, and then the energy of the spin-orbit interaction was obtained. The quantum well profile, as well as the small magnitude of the spin-orbit interaction, allowed us to conclude that the spin-orbit splitting is governed by  the Dresselhaus mechanism.

\end{abstract}


\maketitle

\section{Introduction}

An electron (hole) system in low external magnetic fields or in its absence is usually considered as spin degenerate one.
However, in some cases, namely, when a spin-orbit interaction exists in the system, it turned out that the spin degeneracy can be lifted even in the absence of a magnetic field. Two causes of the spin degeneracy lifting are known. One of them is the crystal structure of the material under study: the lack of the spatial inversion symmetry, which is typical, for example,  for systems with the zinc-blende crystal
structure ($A^{\text{III}}B^{\text{V}}$)~\cite{Dresselhaus}. The second cause is  the structural inversion asymmetry of the quantum well~\cite{Bychkov_Rashba,Bychkov_1984}. The spin-orbit interaction is commonly studied experimentally by measuring the Shubnikov-de Haas (SdH) oscillations. Since the spin-orbit interaction induces formation of two subbands with different spin directions at each wavevector and different concentrations, as well as because the magnitude of the spin-orbit splitting is small, in transport experiments this effect manifests itself in  a beating of the oscillations. By now, beating of the Shubnikov - de Haas oscillations have been observed and extensively studied in many electronic two-dimensional systems. These are InAs/GaSb, InAs/AlSb~\cite{Rowe,Luo}, Ga$_x$In$_{1-x}$As/InP~\cite{Engels,Guzenko,Herzog_2017}, Al$_x$Ga$_{1-x}$N/GaN\cite{Lo}, In$_{0.53}$Ga$_{0.47}$As/In$_{0.52}$Al$_{0.48}$As~\cite{Das,Nitta,Dorozhkin_PRB41}, HgTe~\cite{Minkov_PRB101,MINKOV201995,Gui,Minkov_PRB93}.
The objects of studies in the articles referred to above were mainly heterostructures or one-side doped quantum wells. As for two-dimensional hole systems, aside from work~\cite{Dorozhkin_JETPL}, performed on the InGaAs/GaAs heterostructure, the experiments were carried out mainly on p-GaAs/AlGaAs samples~\cite{Stormer,Eisenstein_pGaAs,Lu,HabibPRB69,Habib_2009,Chiu}. In these works, the objects under study were also, for the most part, heterostructures or one-side doped quantum wells. Moreover, two-dimensional structures were formed on GaAs substrates with the (311)A orientation, and the dopants were either Be or Si. Relatively recently high-quality p-GaAs/AlGaAs structures were grown on GaAs substrates oriented along (100) and doped with carbon (C). The mobility in these structures reached a value exceeding 10$^6$~cm$^2$/Vs ~\cite{Manfra}. In these systems studies of spin-orbit interaction have already been performed in~\cite{Grbic,Nichele,Tarquini,Yuan}.
In both the heterostructure investigated in Ref.~\cite{Grbic}  and the one-side doped quantum well studied in Ref.~\cite{Nichele}, the influence of the spin-orbit interaction on SdH oscillations was clearly manifested. However, in a symmetric quantum well with high mobility and a low concentration studied in ~\cite{Tarquini} no spin-orbit interaction was experimentally observed.
In ~\cite{Yuan}, in a symmetrically doped quantum well ($p=2.2 \times 10^{11}$~cm$^{-2}$, $\mu=0.7 \times 10^6$~cm$^{2}$/Vs at 300 mK), beating of the SdH oscillations were observed. However, its Fourier spectrum peak splitting, which is a specific for the existence of two spin-split subbands, was too weak and could not be analyzed to obtain such characteristics as the effective mass in each subbands.

In this work, we investigate the effect of the spin-orbit interaction on the ac conductance of holes in a 17 nm wide quantum well, symmetrically doped with C. The structure was grown on the (100) GaAs substrate. The measurements were carried out by the Surface Acoustic Waves (SAW) technique in the frequency range 30-300~MHz in magnetic fields up to 18~T, temperatures of 20-300~mK in a linear regime in the wave power, and at $T$=20~mK at various SAW intensities.

\section{Sample parameters and method}

The high-quality samples were multilayer (29 layers) structures grown on a
GaAs (100) substrate.
\resub{The single quantum well in the structure is a 17~nm wide GaAs layer.  It is bounded on both sides by 100~nm layers of Al$_{0.24}$Ga$_{0.76}$As barrier material acting as symmetrical undoped setbacks from the Carbon $\delta$-doped layers on each side. Formed in such a way  quantum well is a square, highly-symmetrical one.}
The quantum well is located at the depth 210 nm below the surface of the sample. It has the hole concentration $p = 1.2 \times 10^{11}$~cm$^{-2}$ and mobility $\mu = 1.8 \times  10^6$~cm$^2$/Vs at 300~mK.

In our studies we utilize the Surface Acoustic Wave technique. In these probeless acoustic experiments a SAW propagates along a surface of a piezoelectric lithium niobate delay line on either edge of which interdigital transducers are placed to excite and detect the wave. The structure under study is pinned down on the surface of the LiNbO$_3$ crystal by means of springs. The electric field accompanying the SAW penetrates into the 2D channel. This ac field induces electrical currents in 2DHG which, in turn, cause Joule losses. As a result of the coupling of the SAW electric field with charge carriers in the quantum well, the SAW attenuation $\Gamma$ and its velocity shift $\Delta v/v$ arise. From simultaneous measurements of $\Gamma$ and $\Delta v/v$ one can determine the complex ac conductance of the 2D structure, $\sigma^{AC}=\sigma_1(\omega)-i\sigma_2(\omega)$. ~\cite{DrichkoPRB2000}

The experiments were carried out in a dilution refrigerator in the magnetic field at $B\leq 18$~T  in the temperature interval 20-300~mK. At the base temperature of 20~mK the magnetic field dependences of the $\Gamma$ and $\Delta v/v$ at various SAW intensity were recorded.  The SAW frequency was changed from 28 to 307~MHz. The measurements in magnetic fields $B<$2~T were done at slow ramping of 0.05~T/min. A Hall probe was used to measure the magnetic field strength precisely.

\section{Experiment results and discussions}

\begin{figure}[h]
\centering
\includegraphics[width=8.5cm]{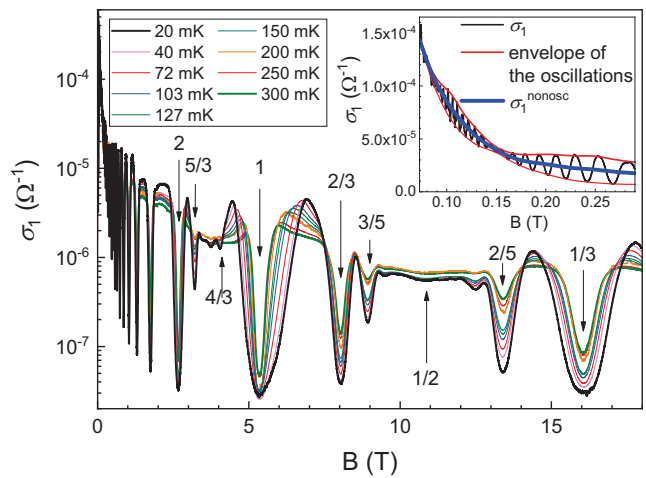}
\caption{(Color online) Magnetic field dependence of $\sigma_1$, at different temperatures and at $f=85$~MHz. The arrows indicate the filling factors
$\nu$. \resub{Inset:  expanded view of $\sigma_1$  for magnetic fields below 0.3~T at 20~mK with the oscillation envelope and its average , i.e., non-oscillating background}.
\label{fig1}}
\end{figure}

Fig.~\ref{fig1} shows the magnetic field dependence of real part of ac conductance $\sigma_1$ at different temperatures at $f\equiv \omega/2\pi =85$~MHz. It shows rich oscillation patterns corresponding not only to the \resub{integer quantum Hall effect (QHE) but also to fractional QHE} regimes, which is typical for a high-quality sample.

At low-field $B<$2~T conductance oscillations undergo beating induced by the spin-orbit interaction~\cite{Stormer,Eisenstein_pGaAs,Lu,HabibPRB69,Habib_2009,Chiu}. In this paper we are focusing on the analysis of this effect and,  demonstrating only the real part of the ac conductance $\sigma_1$ because in this magnetic field $\sigma_2 << \sigma_1$.

\begin{figure}[t]
\centering
\includegraphics[width=8.5cm]{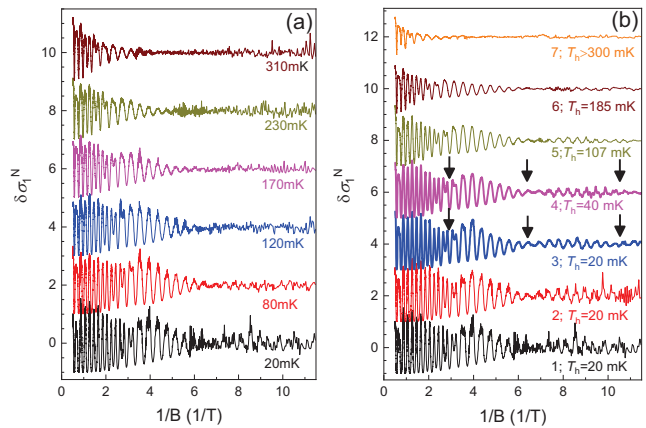}
\caption{(Color online) (a) Dependences of  $\delta \sigma_1^{\text{N}}$ on $1/B$ as
varied with temperature at the SAW intensity of $6\times10^{-10}$~W/cm, and (b) as
varied with the SAW
powers at $T$=20~mK: 1 - 6$\times$10$^{-10}$~W/cm, 2 -
1.9$\times$10$^{-9}$~W/cm, 3 -  6$\times$10$^{-9}$~W/cm, 4 -
1.9$\times$10$^{-8}$~W/cm, 5 - 6$\times$10$^{-8}$~W/cm, 6 -
1.9$\times$10$^{-7}$~W/cm, 7 -
1.9$\times$10$^{-6}$~W/cm. $f$=30~MHz. \resub{The hole temperature $T_h$ is shown for each of curves. See the $T_h$ definition in the text and in Fig. 3.} Traces are offset vertically for clarity. Arrows indicate the oscillations' beating nodes.
 \label{fig2}}
\end{figure}

For this purpose we constructed a dependence of the normalized oscillating part of the conductance $\delta \sigma_1^{\text{N}} = (\sigma_1- \sigma_1^{\text{nonosc}})/\sigma_1^{\text{nonosc}}$ on the reversed
magnetic field 1/$B$, where $\sigma_1^{\text{nonosc}}$ is \resub{monotonically} varying non-oscillating background.
\resub{$\sigma_1^{\text{nonosc}}$ was determined as the average value of the oscillation
envelope in the way illustrated in the inset of Fig.~\ref{fig1}.}
Results of such constructions at various temperatures from 20~mK to
300~mK are presented in Fig.~\ref{fig2}(a). The dependences
$\delta \sigma_1^{\text{N}}$ at $T$=20~mK related to several SAW
intensities
are plotted in Fig.~\ref{fig2}(b), where the
intensity of the SAW electric field introduced into the sample ranges from
$6 \times 10^{-10}$~W/cm to $1.9 \times 10^{-6}$~W/cm.

\begin{figure}[h]
\centering
\includegraphics[width=8.5cm]{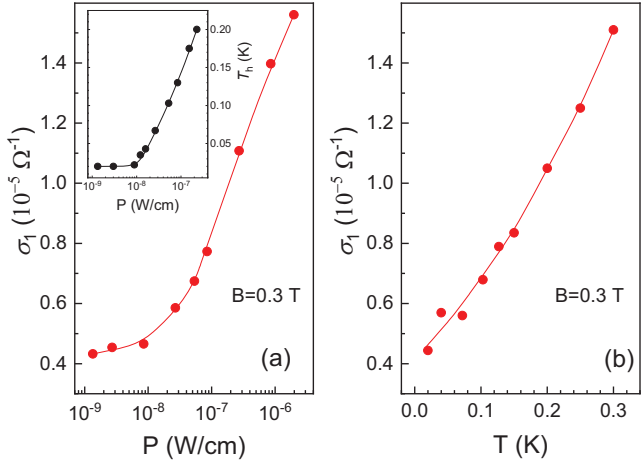}
\caption{(Color online)
Dependences of $\sigma_1$ on the SAW power $P$ at $T=20$~mK
(left panel) and sample temperature $T$ (right panel) at minimal SAW power, both at $B$=0.3~T. $f=85$~MHz. Inset: Dependence of the hole temperature $T_h$ on the SAW intensity. Lines are guides to the eye.
\label{fig3}}
\end{figure}

As seen in Fig.~\ref{fig2}, the conductance undergoes beating with characteristic nodes marked by arrows. The pattern of beating and nodes is much more pronounced on the dependences $\delta \sigma_1^{\text{N}} (B)$ acquired at different SAW intensities (Fig.~\ref{fig2}(b)) rather than at different temperatures (Fig.~\ref{fig2}(a)). This is due to the small value of  the signal-to-noise ratio at the minimum SAW power, at which our study of the temperature influence on the conductivity was carried out.
With  increase in the SAW intensity, the signal-to-noise ratio in the measurements of $\Gamma$ and $\Delta v/v$ increases, but at the same time high intensity  can set the system in a nonlinear regime. To check signal linearity, we plotted the dependence of the conductance $\sigma_1$ in the magnetic field of 0.3~T on the SAW intensity at $T$=20~mK (see Fig.\ref{fig3}(a))  and on temperature at the minimal intensity (see Fig.~\ref{fig3}(b)). Figure 3 evidences that the dependences $\sigma_1 (T)$ and $\sigma_1 (P)$ are similar. Such similarity is typical for heating of current carriers by the SAW electric field. By comparing $\sigma_1 (T)$ and $\sigma_1 (P)$, the dependence of the hole temperature $T_h$ on the SAW intensity was determined and is shown in the inset of Fig.~\ref{fig3}(a). It can be seen from the figure that at $P \leq$10$^{-8}$~W/cm the system is still in the linear regime, i. e., curves 1, 2, and 3 in Fig.~\ref{fig2}(b) correspond to $T_h$=20~mK. However, for the 4th curve, $T_h$ is equal to 40~mK, for the 5th curve $T_h$=107~mK, for the 6th curve $T_h$=185~mK, and for the 7th curve $T_h$ exceeds 300~mK. \resub{The values of $T_h$ are shown for each of curves of Fig.\ref{fig2}(b).}

To clarify the beating structure we carried out the fast Fourier transform (FFT) procedure over the experimental data.

\begin{figure}[t]
\centering
\includegraphics[width=8.5cm]{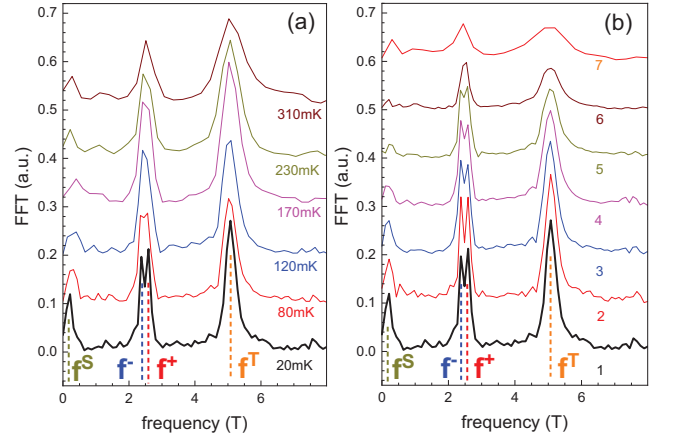}
\caption{(Color online) (a) The
Fourier spectrum of the oscillations in the range $B<$2~T for different temperatures
in linear regime, and (b) for various SAW
intensities at $T$=20~mK: 1 - 6$\times$10$^{-10}$~W/cm, 2 -
1.9$\times$10$^{-9}$~W/cm, 3 -  6$\times$10$^{-9}$~W/cm, 4 -
1.9$\times$10$^{-8}$~W/cm, 5 - 6$\times$10$^{-8}$~W/cm, 6 -
1.9$\times$10$^{-7}$~W/cm, 7 -
1.9$\times$10$^{-6}$~W/cm. Traces are offset vertically for clarity.
 \label{fig4}}
\end{figure}

Fig.~\ref{fig4} shows the Fourier spectra of the conductance for magnetic field $B<$2~T for different temperatures in linear regime (a) and for various SAW electric field intensities at 20~mK (b). The Fourier
spectra exhibit several components. We see here clearly, especially at low temperatures, a split FFT peak with the FFT frequencies f$^- \cong$2.4~T and f$^+ \cong$2.6~T resulting in appearance of the both the slow oscillation FFT peak at f$^{\text{S}} \cong$0.2~T and the sum FFT peak at f$^{\text{T}} \cong$5~T. The frequencies can be converted into the densities
of the lower  and higher populated subbands: $p^- \cong 5.8 \times 10^{10}$~cm$^{-2}$ and $p^+ \cong 6.3 \times 10^{10}$~cm$^{-2}$, respectively. Therefore, $\Delta p \equiv p^{\text{S}} = p^+ - p^- \cong 5 \times 10^{9}$~cm$^{-2}$ and $\Delta p / (p^+ + p^-) \approx 4\%$.

This all confirms that the slow oscillations
appear due to the intersubband transitions.
Moreover, it is supported by a robustness of the f$^{\text{S}}$ FFT peak to the  temperature and SAW power increase, even when the FFT f$^+$ and f$^-$ split is not already observed.

\begin{figure}[t]
\centering
\includegraphics[width=8.5cm]{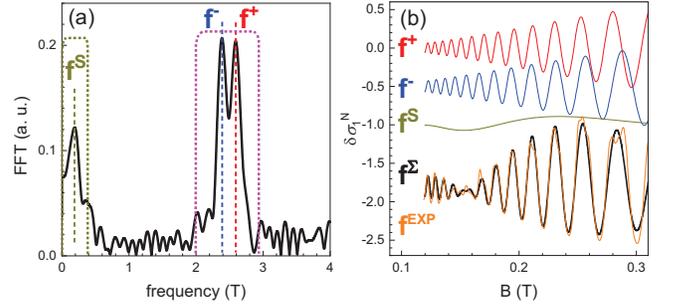}
\caption{(Color online) (a) The
Fourier spectrum of the oscillations in the range $B<$1~T and the bandpass filters (the dotted
lines) used to isolate the fast and slow components. (b) The oscillating components of $\delta \sigma_1^{\text{N}}$ in upper f$^+$ and lower f$^-$ subbands, f$^{\text{S}}$ is the slow oscillation component. \resub{Also shown is the sum curve f$^{\Sigma}$= f$^+$ + f$^-$ +f$^{\text{S}}$ (black) as compared with the experimental one f$^{\text{EXP}}$ (orange). } Traces in (b) are offset vertically for clarity.
 \label{fig5}}
\end{figure}

Now we analyze the behavior of different components of the FFT spectrum of the conductance. First, we used a
bandpass filter, shown in Fig.~\ref{fig5}(a) as an olive dotted line,  to isolate the slow component from fast ones. Then, applying the
inverse fast Fourier transformation we obtained the slow oscillations of the frequency f$^{\text{S}}$, which is shown in Fig.~\ref{fig5}(b).

Secondly, we should isolate from each other the two fast oscillations corresponding to upper f$^+$ and lower f$^-$ subbands. Since f$^+$ and f$^-$ are almost equal bandpass filtering has an ambiguity when constructing the corresponding bandpass filter. That is why the direct inverse fast Fourier transformation would not be precise.
Therefore, we used a specially developed software provided to us by O.~E.~Rut. The software algorithm is described in Ref.~\cite{Minkov_PRB101}. It carries out the bandpass filtering of the \textit{splitted} FFT peak with a filter shown in Fig.~\ref{fig5}(a) as a pink dotted line. Then, the inverse fast Fourier transformation gives a sum
of fast oscillations with frequencies f$^+$ and f$^-$. Finally this sum is fitted to the sum of two Lifshits-Kosevich~\cite{Lifshits_Kosevich}
 terms presented in  Eq.~(6) of Ref.~\cite{Minkov_PRB101}:
\begin{eqnarray}
  \label{LS}
&& \delta \sigma_1^{\text{N}}=
\sum^2_{i=1}\beta_i \exp(-\frac{2\pi\gamma_i}{\hbar \omega_c^i})
D(X_T^i) \times \,   \nonumber \\
&&
\times \cos \left(\frac{2 \pi f_i}{B} + \phi_i\right),
\end{eqnarray}
where $D (X_T^i)= X_T^i / \sinh (X_T^i)$, $X_T^i=2\pi^2 k_B T / \hbar \omega_c^i$.
The fitting parameters $f_{1,2}$, $\gamma_{1,2}$, $\beta_{1,2}$, $\phi_{1,2}$ are described in Ref.~\cite{Minkov_PRB101} and references therein.
Fast oscillations f$^+$ and f$^-$ reconstructed by using the software mentioned above are shown in Fig.~\ref{fig5}(b). \resub{The quality of such fitting is also illustrated in this figure by  comparing  the superposition curve f$^{\Sigma}$=f$^+$ + f$^-$ + f$^{\text{S}}$ with the experimental one f$^{\text{EXP}}$.}

Such procedure was performed for all temperatures in the field region, where beating were observed. It allowed us to build the dependences of the amplitude of oscillations on temperature in different magnetic fields.

In a 2D system with one subband the carrier effective mass $m^*$ is usually determined from
 the temperature dependence of the amplitude of the SdH-type
oscillations by fitting
it to the factor $D (X_T)$ with the effective mass $m^*$ as the fitting
parameter.

\begin{figure}[t]
\centering
\includegraphics[width=8.5cm]{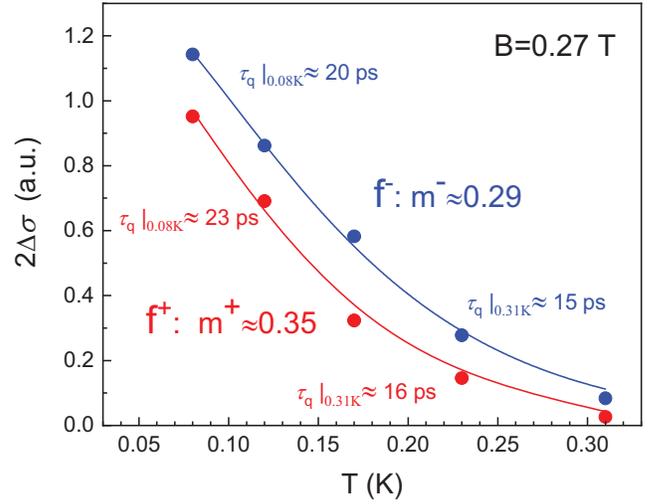}
\caption{(Color online) The
temperature dependence of 2$\Delta \sigma$ for
components f$^+$ and f$^-$. The solid lines are fittings using the  thermal damping term of the Lifshitz-Kosevich formula.
 \label{fig6}}
\end{figure}

In our case, we can independently determine $m^*$ from temperature dependences of amplitudes as we isolate from each other the fast oscillations corresponding to upper f$^+$ and lower f$^-$ subbands.  Fig.~\ref{fig6} shows dependence of the oscillation amplitudes 2$\Delta \sigma$ of each component for magnetic field 0.27~T.  Fits
of the oscillation amplitudes to $D (X_T)$ provide us with $m^*$ for each subband.

\begin{figure}[t]
\centering
\includegraphics[width=8.5cm]{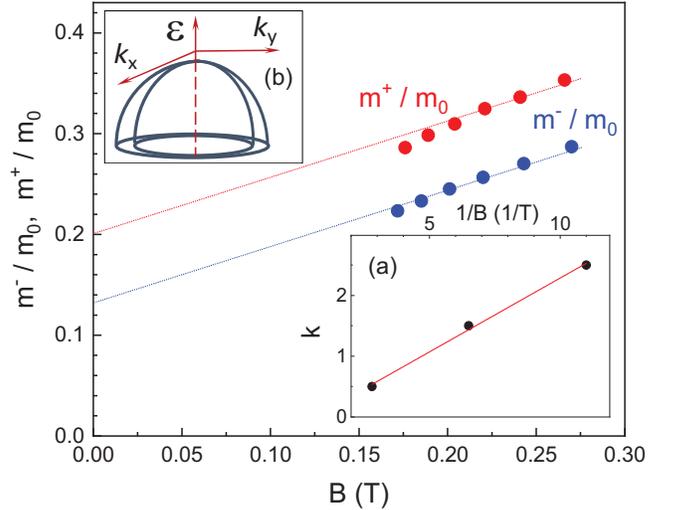}
\caption{(Color online)
The dependence of $m^+ / m_0$ and $m^- / m_0$ for components f$^+$ and f$^-$ on the magnetic field, $f$=30~MHz. Lines are  extrapolations of dependencies $m^+ / m_0 (B)$ and $m^- / m_0 (B)$ to $B$=0.
Inset (a): dependence of the oscillations beating nodes number k on 1/$B$.
\resub{Inset (b): Sketch of the dispersions of holes in a
2DHG in the presence of spin-orbit interaction for the Dresselhaus mechanism}.
 \label{fig7}}
\end{figure}

Fig.~\ref{fig7}
shows $m^*$ values as a function of magnetic field $B$ for each of the  spin-split subbands. $m^+$ and $m^-$ nearly linearly increase with
$B$. Linear extrapolations of the data to
$B=$0 shown by lines in Fig.~\ref{fig7}  suggest \resub{$m^+ / m_0$=0.20$\pm$0.01 and $m^- / m_0$=0.12$\pm$0.01}, where $m_0$ is the mass of free electrons.

The quantum scattering times were also obtained from the oscillations' amplitudes: at $B$=0.27~T the value are $\tau_q \approx $~1.6$\times$10$^{-11}$s for upper subband and $\tau_q \approx $~1.5$\times$10$^{-11}$s for lower subband at $T$=0.31~K. The transport relaxation time \resub{$\tau_{tr} \approx $1.62$\times$10$^{-10}$s} found from zero-field mobility of $\mu = 1.76 \times  10^6$~cm$^2$/Vs at 0.31~K with value of the effective mass $m^*$=(0.16$\pm$0.04)$m_0$ averaged between $m^+$ and $m^-$ at $B$=0. The ratio $\tau_{tr}/\tau_q \approx$12
indicates that a long-range scattering potential is dominant.

Magnetic field dependences of the effective mass for spin-split subbands formed as the  result of the spin-orbit interaction are reported in all studies of this effect~\cite{Lu,HabibPRB69,Habib_2009,Chiu,Nichele,Tarquini}. A possible theoretical explanation of the origin of this dependence was proposed in Ref.~\cite{SimionLyanda-Geller}. However, the quantum well width, as well as the sample parameters used in the calculations published in Ref.~\cite{SimionLyanda-Geller}, differ from those studied in our work. It makes a comparison with the theory of Ref.~\cite{SimionLyanda-Geller} impossible.

Now, since we have estimated the values of effective masses at $B$=0 it is possible to calculate the energy of the spin-orbit splitting $\Delta_{\text{SO}}$. This calculation can be done by two methods.

In the first method, knowing the difference in concentrations in the subbands split by the spin-orbit interaction $\Delta p \approx p^+ - p^- \approx $ ~5$\times$10$^9$~cm$^{-2}$, we use  the following  formula:
\begin{equation}
\label{DSO}
\Delta_{\text{SO}} =  \Delta p 2 \pi \hbar^2 /m^* = \resub{(0.16 \pm 0.04)~\text{meV}},
\end{equation}
\resub{which is valid in our case since $\Delta p \ll p$.}

In the second method, since it is possible to determine the magnetic field position of oscillations beating nodes (see Fig.~\ref{fig3}), and the
modulation of the SdH-type oscillations induced by the spin-orbit interaction is realized according to the law
$\sigma \propto \cos (\pi\Delta_{\text{SO}}/ \hbar \omega_c)$,
then $\sigma = 0$ at \resub{(in CGS units)} k=$\Delta_{\text{SO}}/ \hbar \omega_c= \Delta_{\text{SO}} m^* c / e \hbar B$ = 1/2, 3/2, 5/2. From the slope of the
dependence of k on 1/$B$ shown in inset of Fig.~\ref{fig7}, one can determine \resub{$\Delta_{\text{SO}}$=(0.2$\pm$0.05)~meV}.
In both of these methods, as well as in the mobility calculations above, we used averaged \resub{$m^*$=(0.16$\pm$0.04)$m_0$}.
\resub{As we see the values of $\Delta_{\text{SO}}$ determined by different methods are close.}

As was mentioned in the introduction, the scientists who  studied the spin-orbit interaction in two dimensional structures in most cases dealt either with heterostructures or with one-side doped quantum wells. Therefore, almost all of them observed the effect related to structural inversion asymmetry
(Rashba splitting), which was much stronger than the bulk inversion asymmetry effect (Dresselhaus splitting) expected in symmetrically doped quantum wells.
\resub{Also it should be mentioned that the Dresselhaus spin-orbit interaction was studied in $n$-GaAs/AlGaAs using the weak antilocalization in magnetoresistance effect and the energy $\Delta_{\text{SO}}$=0.2~meV was obtained~\cite{Desrat}.}
Comparing our results with the values of the Rashba energy splitting acquired by these authors on various two dimensional systems: $\Delta_{\text{SO}}$=30~meV~\cite{Gui}, 9~meV~\cite{Lo}, 5.5~meV~\cite{Nitta}, 1.41~meV~\cite{Stormer}, we established that theirs were significantly higher than the value obtained by us: \resub{$\Delta_{\text{SO}}$=(0.16$\pm$0.04)~meV and $\Delta_{\text{SO}}$=(0.2$\pm$0.05)~meV}.
 Since we investigated the spin-orbit interaction in a symmetrically doped square quantum well and obtained a very small value of the spin-orbit splitting, there is a reason to believe that the origin of this effect intrinsically comes from the crystal structure of the material under study (i.e., from the absence of an inversion center), which is typical for  zinc-blende ($A^{\text{III}}B^{\text{V}}$) structures (Dresselhaus mechanism).
\resub{Shown in the inset (b) of Fig.~\ref{fig7} for simplified visualization is the holes spectrum with a presence of the Dresselhaus spin-orbit interaction mechanism. The spectrum consists of two spin-subbands, giving oscillations with close frequencies resulting in a beating.}

It is worth noting that determined in the present paper values of the effective masses, quantum relaxation times and energy of the spin-orbit interaction do not depend on frequency in the studied range of 30-300~MHz.

In conclusion, namely using of the high-quality structure $p$-GaAs/AlGaAs (with $p = 1.2 \times 10^{11}$~cm$^{-2}$) with a square symmetrically doped quantum well and carrying out measurements of the SdH-type oscillations in the low temperature region 20~mK~$<T <$300~mK gave us the possibility to observe for these structures the spin-orbit interaction which is originated from the Dresselhaus effect.

\paragraph*{Acknowledgments}
The authors would like to thank O.E. Rut for providing us with the FFT components separation program, G.M. Minkov for fruitful consultations and L.E. Golub for useful discussions and careful reading of the manuscript.
The authors are thankful to E. Green, P. Nowell, and L. Jiao for technical assistance. Support from
 the Russian Foundation for Basic Research (project
19-02-00124)
is gratefully acknowledged. The National High Magnetic Field Laboratory is supported by National Science Foundation through NSF/DMR-1644779 and the State of Florida. This research is funded in part by the Gordon and Betty Moore Foundation’s EPiQS Initiative, Grant GBMF9615 to L. N. Pfeiffer, and by the National Science Foundation MRSEC grant DMR 1420541.


%

\end{document}